\documentclass[prb,twocolumn,showpacs]{revtex4}
\usepackage{amsfonts,amsmath,mathrsfs,epsfig,amsbsy,bm,verbatim}

\newcommand{\ket}[1]{\left\vert {#1} \right\rangle}
\newcommand{\expect}[1]{\langle {#1} \rangle}

\begin{document}
\title{A number conserving theory for topologically protected degeneracy in one-dimensional fermions}
\author{Jay D. Sau $^1$}
\thanks{Present address: Department of Physics, Harvard University, Cambridge, MA 02138, USA.}
\author{B. I. Halperin $^{2}$}
\author{K. Flensberg $^3$}
\author{S. Das Sarma $^1$}

\affiliation{$^1$Condensed Matter Theory Center and Joint Quantum Institute, Department
of Physics, University of Maryland, College Park, Maryland 20742-4111, USA.
$^2$Department
of Physics, Harvard University, Cambridge, Massachusetts, USA.
$^3$Niels Bohr Institute and Nano-Science Center, University
 of Copenhagen,
Universitetsparken 5, DK-2100 Copenhagen, Denmark.}

\begin{abstract}
Semiconducting nanowires in proximity to superconductors are among 
promising candidates to search for Majorana fermions and topologically
 protected degeneracies which may ultimately be used as building blocks for
 topological quantum computers. The prediction of neutral Majorana
 fermions in the proximity-induced superconducting systems ignores
 number-conservation and thus leaves open the conceptual question of how 
a topological degeneracy that is robust to all local perturbations 
arises in a number-conserving system.
 In this work, we study how local attractive interactions
generate  a topological ground-state near-degeneracy
 in a quasi one-dimensional
 superfluid using bosonization of the fermions.
The local attractive interactions opens a topological 
 quasiparticle gap in the odd channel wires
 (with more than one channel) with end Majorana modes associated
 with a topological near-degeneracy.
 We explicitly study the robustness of the topological degeneracy 
to local perturbations and find that such local perturbations result 
in quantum phase slips which  split of the 
topological degeneracy by an amount that  does not decrease
 exponentially with the
 length of the wire, but still decreases rapidly if
 the number of channels is large.
 Therefore a bulk superconductor with a 
large number of channels is crucial for true topological degeneracy. 
\end{abstract}

\maketitle

\section{Introduction}
In the past few years, topological superconductors have
become promising candidates for realizing Majorana fermions (MFs) 
 \cite{volovik,kopnin,sumanta,read_green,Zhang,fu_prl'08,schnyder,Sato-Fujimoto,sau,long-PRB},
 which are of interest both for fundamental reasons
 as a new type of particle with non-Abelian statistics 
and also their potential application in topological quantum computation
 (TQC)\cite{nayak_RevModPhys'08,Wilczek-3,levi,science}.
One of the simplest such topological superconducting (TS) 
systems supporting MFs consists of a semiconductor nanowire in a  
magnetic field together with spin-orbit coupling and superconductivity 
\cite{roman,oreg,long-PRB}
through the appropriate tuning of the semiconductor chemical 
potential or equivalently, the carrier density. 
The superconducting pair-potential is induced in the nanowire by
 the proximity effect from an $s$-wave
 superconductor in contact with the nanowire. It has been shown that
 such a nanowire can be driven into 
a TS phase with MF end modes   \cite{kitaev,long-PRB,roman,oreg,franz,hole-doped}.

\begin{figure}
\centering
\includegraphics[scale=0.3,angle=0]{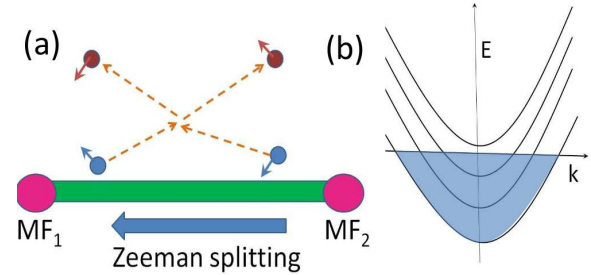}
\caption{(Color online) (a) Combination of Zeeman splitting, spin-orbit coupling
 and weak attractive $s$-wave 
Feshbach resonance can induce end MFs in multi-channel 
nanowire (shown as green rectangle). Spin-orbit coupling and Zeeman splitting ensures almost 
helical fermions at the fermi surface
 (shown as discs with arrows representing spin).
 Attractive $s$-wave interaction leads to the scattering of pairs of 
fermions with zero momentum (Cooper pairs)
 from one channel (represented by blue color) 
to another (represented in red).
(b) Schematic band-structure of multi-channel nanowire with all
 degeneracies lifted by Zeeman splitting. Filling corresponding to
 odd number of bands is in the topological phase.
 }\label{Fig1}
\end{figure}

The MFs in the TS phase of the semiconducting nanowire are localized
 at the ends of the wire and can be 
detected as zero-bias conductance peaks in the tunneling spectrum at
 the ends \cite{kitaev,yakovenko,long-PRB}.  The end MFs are associated 
with  self-adjoint
 operators $\gamma_1$ and $\gamma_2$  ($\gamma_i^{\dagger}=\gamma_i$) 
and in this sense MFs are their own anti-particle.
 MFs are of interest for TQC because despite having no internal
 degrees of freedom individually, a pair of
 MFs has two distinct possible states (fusion channels).
 These states are energetically degenerate to a degree exponential
 in the separation
 of the MFs, and correspond to states where the combined
 fermionic state $f^\dagger=\frac{\gamma_1+i\gamma_2}{\sqrt{2}}$ 
is occupied (particle fusion channel) or empty (vacuum fusion channel).
 These channels correspond to even and odd total number of fermions in
 the system. The MFs in a system of nanowires have a non-Abelian
 statistics under exchange of positions of the MFs \cite{alicea1,david1,david}.
 The result of
 exchange of MFs in the nanowire system, which can be determined by
 measuring the  channel (particle or hole) each pair of MFs fuses into,
 depends on the order in which the exchanges are performed. This
 non-Abelian statistics forms the basis of topologically protected
 operations using MFs. Most studies of MFs in TS have so far been
 restricted to non number-conserving cases either via mean-field BCS 
 studies or for systems where the superconductivity is induced by
 the proximity effect.
 Within these formalisms the MF is necessarily neutral. However 
tunneling of electrons into MF modes results in a transition of the
 nanowire between the quasiparticle and vacuum fusion channels 
which have different total electron numbers. Therefore a proper
 description of tunneling through a neutral MF mode must involve 
charge fluctuations where the initially localized electronic charge
 delocalizes over the whole wire. A recent study by Leggett 
\cite{leggett} discusses the issue of manifestly including 
number conservation and constructs a number-conserving 
version of a $p$-wave topological 
superconducting wire Hamiltonian\cite{kitaev} with topological 
degeneracy for a particular set of parameters. 

In this paper, we analyze the consequences of number conservation
 by comparing several  models in which
 we have one or more coupled  nanowires in the TS state
 where superconductivity is generated by 
intrinsic attractive interactions. The calculations in this paper 
are performed within the bosonization technique \cite{haldane,giamarchi},
so that any recourse to mean-field BCS-type approximations that 
break number conservation is avoided.
 Each wire is supposed to  contain an odd number $N_{ch}$
 of non-degenerate, spin-orbit split,  transverse channels. 
In our calculations, we assume that the wires may have an intrinsic
 superconductivity, induced by attractive two-body interactions of
 finite range.
We will also comment on the comparison of this scenario to the 
case where superconductivity is induced by the proximity effect 
from a neighboring bulk superconductor.
As such our work will also apply to the one-dimensional topological 
insulator nanowire system which also has an odd number of conducting 
channels.\cite{franz}  Our focus will be on enumerating the
 low energy states of the wire network,
 associated with MF modes at the ends or at junctions
 between wires, and on estimating the residual energy
 splittings between the low energy states which occur when
 the wires are long but not infinite in length.
  In particular, we will want to distinguish between energy splittings that fall off exponentially in the
 length of the wire segments, and those which have a weaker dependence on length.  Our analysis will use
 bosonization techniques appropriate for one-dimensional fermions, together with renormalization group and scaling analyses where needed.\cite{haldane,giamarchi}
The bosonization technique has already been applied in the context of TS nanowires to understand the effect of 
interactions \cite{alicea} in non-number conserving
 proximity-induced superconducting systems.\cite{loss,roman_fisher}

We note that a few-channel nanowire \cite{roman-tudor} with local attractive interactions can be potentially be realized in cold-atom systems with quasi-one-dimensional confinement, together with Zeeman spitting, spin-orbit coupling \cite{spielman} and attractive s-wave Feshbach interactions \cite{Zhang-pwave}.  However, the primary motivation for this paper is to explore the conceptual issues related to conservation of particle number\cite{leggett}
and how topological degeneracy may be described in such a framework.

In the case of proximity-induced superconductivity, it has typically been assumed that the bulk superconductor is infinitely large, so that the phase of the superconducting order parameter may be  perfectly well defined, and one does not need to worry about conservation of total charge.  Moreover, it has generally been assumed that bulk superconductors attached to different wires are strongly coupled to each other, so that they all have the same superconducting phase, and one does not have to worry about effects of phase slips.  In our discussions below, we shall relax this last assumption and comment on the situation when phase slips do occur in the connections between different wires, either due to thermal activation or due to quantum-mechanical tunneling.
For the case of nanowires with intrinsic superconductivity, we shall primarily be interested in networks of several wires, which are coupled to each other through  tunnel  barriers over large portions of their lengths.  Results will depend on the number of wires involved and the ways in which they are coupled.
\section{Outline and Results}
We first consider the case of single isolated wire. To clarify the role of number conservation, we start in Sec.~\ref{number} 
by discussing the differences between the degeneracies of proximity-induced TS wires and intrinsic TS wires.
Following this discussion, it becomes clear that for an isolated wire,
 the size of the energy gap will not decrease exponentially with the length  $L$ of the wire, but will
 generally be inversely proportional to $L$.  This is a direct consequence of particle conservation in the isolated wire.
Similar splitting has been predicted for MFs on superconducting grains. \cite{majorana_teleportation}  

In order to find a situation where the tunneling gap and the lowest particle-hole excitation fall off faster than $1/L$,
  we must consider at least two coupled wires. Before discussing multi-wire systems, we start in Sec.~\ref{secii} by 
reviewing how to determine the properties of weakly interacting multi-channel wires using bosonization.
  In Sec.~\ref{seciii}, by computing the scaling dimensions of the Cooper-pair scattering terms, 
we find that for  an infinite wire, with $N_{ch} >1$, for weak interactions which are attractive in all channels
 (see more precise definition below) the system develops an energy gap for single electron tunneling at points away
 from the ends of the wire, along with quasi-long-range power-law correlations in the superconducting order parameter . \cite{giamarchi}
As we will discuss in Sec.~\ref{seciii}, despite the absence of a global superconducting phase,
 bosonization allows us to assign a local superconducting phase $2\theta_a(x)$ to each of the $N_{ch}$ channels. 
In this phase,  the values of $\theta_a(x)$ are locked to each other up to a multiple of $\pi$.
 Thus, the phases of each channel relative to the local common superconducting phase  forms a $Z_2$ order similar to the Ising model.
 In fact, at this level, bosonization maps the multi-channel wire into $N_{ch}$ copies 
of spinless $p$-wave TS wires \cite{kitaev} coupled to a global phase field, each of which can be mapped to a transverse field Ising model  by a 
non-local Jordan-Wigner transformation.  \cite{auerbach}
However, this $Z_2$ phase representation does not manifestly respect the particle number conservation
 symmetry of the intrinsically number conserving 
interacting TS Hamiltonian. 
This is remedied in Sec.~\ref{hilbert} by identifying the appropriate total number subspace which is preserved by the Hamiltonian. The restriction of the 
Hilbert space reduces the topological degeneracy from $2^{N_{ch}}$ to $2^{N_{ch}-1}$.

 The degeneracy associated with $Z_2$ order parameters for each channel is topologically protected only if it is 
robust to all local perturbations in the fermionic representation. 
 Bosonization is a non-local transformation, similar to the Jordan-Wigner transformation, and 
therefore, the local $Z_2$ order associated with each channel in the bosonic representation is a non-local
 and potentially topological order parameter 
in the original 
fermionic representation. However, the topological robustness to local fermionic perturbations is something that must be explicitly checked.
Since local fermionic tunneling operators are gapped in the middle of the wire, the ends of the wires are expected to be 
more susceptible to fermionic perturbations.
By considering all such local fermion terms at the ends of the wires, in the bosonic representation in Sec.~\ref{secv},
 we show explicitly in Sec.~\ref{sectop}, that the topological degeneracy is robust to perturbations at the ends of the wire only for wires with an odd number
 of channels $N_{ch}$. In the absence of the number conservation constraint, the end scattering reduces the degeneracy of an isolated wire from 
$2^{N_{ch}}$ to $2^{N_{ch}(mod\, 2)}$. This is similar to the familiar splitting of pairs of MFs in the proximity-induced case.\cite{kitaev}
 However,  since in this paper we are interested in finding exponential in system size splittings of topological 
degeneracies, we derive 
this result entirely in the bosonic representation without recourse to refermionization. 
Thus, topological degeneracy can only survive for $N_{ch}$ odd. Of course, for the number conserving case,
 the degeneracy is completely eliminated and 
one must use a bundle of wires to obtain a degeneracy.
 
The robustness to scattering in the middle of the is a more subtle issue that is clarified later.
 For a wire with ends, the topological degeneracy manifests itself as states with energies inside the bulk gap,
 which can be accessed by tunneling of an electron in the vicinity of either end.
  The amplitude for tunneling to these states falls off exponentially as the tunneling point moves away from either end. 
 For a wire with an odd number of channels, the energy gap for tunneling near the ends of a wire will decrease as the wire length
 increases, and the relevant states will be reminiscent of the Majorana end states for a wire with proximity-induced superconductivity.

As will become clear from both the general considerations in Sec.~\ref{number} and also the detailed bosonization treatment in the 
following sections, in order to find a situation where the tunneling gap and the lowest particle-hole excitation fall off faster that $1/L$,
  we must consider at least two coupled wires as discussed in Sec.~\ref{secxi}.  The simplest case is to consider $N_w$ parallel wires of length $L$,
  each with an odd number of  channels, which are coupled to each other only over the middle third of their length.  (Fig.~\ref{Fig2}(b))
 If the tunnel coupling between the wires is sufficiently strong, we expect that the number of electrons in each separate wire will fluctuate by a
 large amount, (with root-mean-square fluctuation proportional to $L^{-1/2}$). Then the fermion parity in any particular wire will have little
 effect on the energy, as long as the total fermion parity is fixed.  Thus the energy difference between the lowest  state in which say wires
 1 and 2 have both have even parity, or they both have odd parity, could be quite small.  Indeed, we find that under ideal conditions, where the
 tunneling between the wires turns on adiabatically as a function of the distance along the wires, the energy splitting between lowest energy
 states of the system can be exponentially small  in $L$. This is shown in Sec.~\ref{exponential} by analyzing the role of instanton tunneling 
terms where the global phase of an entire channel can jump by $\pi$ from one energy minimum to another.
The requirement of strong coupling between two wires is achieved when the Josephson coupling energy between the wires is large compared to the
 charging energy of either wire. This is easily achieved when $L$ is large because the Josephson coupling strength increases proportional to the
 length of the coupling region, while the charging energies fall off as $1/L$.

In Sec.~\ref{secix}, we address the robustness of the topological degeneracy to local fermionic perturbations in the middle of the wire.
Here, the situation becomes more complicated
since back-scattering generates kinks in the superconducting phases of various channels which are 
identical to impurity generated quantum phase-slips.  In the case where the number of wires $N_w$ in the bundle is even, we find that backscattering,
 {\em even in the center of the wire bundle}, is capable of destroying the exponential $L$-dependence of the energy splitting. 
 When the number of wires is odd, this does not happen, but any backscattering that is present due to violations  of momentum conservation
 {\em near  a junction between two wires} will still lead to destruction of the exponential dependence of the energy spitting. (Fig.~\ref{Fig2}(b))
 Therefore, topological degeneracy that is exponential in the length of the wire can only occur if all wires are in contact with a nanowire with a
 very large number of channels. The different responses of total odd and even number of channel systems to quantum phase slips can be understood in 
terms of the fractional Josephson effect shown by TS wires with odd number of channels.\cite{kitaev,yakovenko,roman,oreg,fu_prl'08}

In Sec.~\ref{secx}, we find that there is another important difference in the spectra for tunneling into the end of a TS wire which distinguishes
 the proximity-induced superconductor from the isolated wire.  In the proximity-induced case, there will be a finite amplitude for tunneling into the
 zero energy state, even in the limit of $L\to \infty$.  For the isolated wire, the amplitude for tunneling into the lowest energy state at the end
 of a wire falls off as a power of the length $L$.   The missing spectral weight is diverted to states in which varying numbers of low energy
 ``phonons"
( These phonons of course are the collective bosonic modes of the electron system, not of the underlying lattice, with a gapless linear energy dispersion in momentum, arising from the short-range electron-electron interaction.)
 are simultaneously excited. In the limit $L \to \infty$, this becomes a continuum with a spectral weight diverging at zero energy.  

Finally, in Sec.~\ref{secxii}, we show that despite the presence of gapless phonons that broaden the end tunneling peaks associated with MFs,
 it is still possible to manipulate
the topological degenerate space in wire bundles. This is done using a tunneling protocol which changes the tunneling between MFs in such a 
way that the tunneling pins the local superconducting phase around the MFs before introducing tunneling between the MFs.

\section{General consequences of number conservation}\label{number}
To understand the difference between the isolated wire and the wire with proximity-induced superconductivity, recall that the two low-lying states formed from the MF modes at the wire ends correspond to states of opposite fermion parity for the wire.  For the proximity-induced situation, tunneling of Cooper pairs will cause large fluctuations in the number of electrons in a long wire, while preserving the fermion parity.   When the number fluctuations are large compared to unity,  the difference in energy between an even  and odd parity state will become exponentially small in the length of the wire.  By contrast, for an isolated wire,  where the Hamiltonian commutes with the total electron number, any energy eigenstate of the wire will be characterized by a definite particle number $N$.  For small deviations of $N$ from some reference number $N_0$, we may write  the ground state energy as
\begin{equation}
E_0 (N) = E_0(N_0) + a (N-N_0) + b (N-N_0)^2 + ...
\end{equation}
The constant $b$ may be identified with the \textquotedblleft charging energy" for the wire; it is inversely proportional to the compressibility and it  falls off as $1/L$, for a long wire with short-range interactions. The constant $a$ depends on the reference energy for an isolated fermion, or on the chemical potential of a reference reservoir, so it has no intrinsic meaning.  The proper definition of the energy gap for tunneling into the ground state of the wire is one-half the difference in the energies necessary to add or subtract an electron from the wire, which is simply equal to $b$ in the present case. We note that the lowest  energy for a particle-hole excitation in a finite wire will generally be the energy to excite a phonon in the mode with wavelength $L/2$, which also varies as $1/L$ for large $L$, and is of similar magnitude to $b$.  Therefore to achieve true topological degeneracy it will be necessary to move away from the paradigm of an isolated wire as will be done in Sec.~\ref{secxi}

\section{Nanowire Hamiltonian and bosonization}\label{secii}
While an isolated wire cannot show true topological degeneracy,  it is 
instructive to study isolated nanowires before analyzing systems of wires. 
An isolated spin-orbit coupled multi-channel nanowire with Zeeman splitting
 has a band-structure shown in Fig.~\ref{Fig1}. For a system with 
weak interactions, one can linearize the fermionic excitations in the
 vicinity of the Fermi energy $\mu$ 
\begin{equation}
\psi^\dagger_a(x)= e^{i  k_{F,a} x}\psi^\dagger_{a,L}(x)+e^{-i k_{F,a} x}\psi^\dagger_{a,R}(x)
\end{equation} 
where $k_{F,a}$ and $\psi^\dagger_a(x)$ are the fermi wave-vector and 
fermionic operator associated with band $a=1,\dots,N_{ch}$ while $\psi^\dagger_{a,R}(x)$ 
and $\psi^\dagger_{a,L}(x)$ represent fermionic excitations around 
$k\approx \pm k_{F,a}$ respectively \cite{haldane,giamarchi}.  
The linearized non-interacting 
part of the Hamiltonian is written as: 
\begin{align}
&H_0\approx\sum_a \int dx \psi^\dagger_{a,L}(x)(i v_{F,a}\partial_x)\psi_{a,L}(x)\nonumber\\
&-\psi^\dagger_{a,R}(x)(i v_{F,a}\partial_x)\psi_{a,R}(x).
\end{align}
The attractive interaction between the fermions can be written as a product of $4$-fermion operators ($\psi^\dagger\psi^\dagger\psi\psi$)
 and conserves total momentum so that only 2 classes of terms are allowed.
 The first of these classes
\begin{equation}
H_{1}=\int dx \sum_{a,b,r,r'}V_{r,r',a,b}\rho_{a,r}(x)\rho_{b,r'}(x),
\end{equation}
where $\rho_{a,r}(x)=\psi^\dagger_{a,r}(x)\psi_{a,r}(x)$,
 involves the densities of left and right
 movers in each band. The other class is a back-scattering term 
\begin{equation}
H_{2}=\int dx \sum_{a \neq b}V^{(pair)}_{a,b}\psi_{a,L}^\dagger(x)\psi_{a,R}^\dagger(x)\psi_{b,R}(x)\psi_{b,L}(x),
\end{equation}
which may be thought of as transferring Cooper pairs ($\psi_{a,L}^\dagger(x)\psi_{a,R}^\dagger(x)$)from band $b$ to $a$.
The latter term will be shown to lead to a quasiparticle gap of the 
superconducting kind when $H_1$ contains attractive interactions.
In this paper, we analyze the quasi-one dimensional Hamiltonian $H=H_0+H_1+H_2$ using the Abelian bosonization 
identities\cite{haldane,giamarchi,gogolin,schoeller} to represent the fermions in terms of bosonic fields as:
\begin{equation}
\psi_{a R}(x)=\frac{F_a }{\sqrt{2 \Lambda}}e^{i\pi \frac{x}{L}\Delta N_a}e^{i\vartheta_a(x)}\label{bosonization}
\end{equation}
together with $\psi_{a L}(x)=-\psi_{a R}(-x)$ where $\psi_{a R}(x)=\psi_{a R}(x+2 L)$ is assumed to be periodic and defined from $[-L,L]$, $L$ 
is the length of the wire and $\Lambda$ is a short length-scale cut-off scale.
The operator $\Delta N_a$ is the number fluctuation operator that is associated with 
the band $a$ defined as 
\begin{equation}
\Delta N_a=N_a-\frac{k_F L}{2\pi},
\end{equation}
where $N_a$ is the number of electrons in channel $a$, 
and the Klein factors $F_a$ are constructed for this case as 
\begin{equation}
F_a=e^{i\pi\sum_{a'<a}\Delta N_{a'}}
\end{equation}
to ensure canonical anti-commutation relations of the fermion operators \cite{schoeller}.
The bosonic field $\vartheta_a(x)$ has a somewhat involved commutation 
relation \cite{gogolin}, which can be simplified by rearranging $\vartheta_a(x)$ into a pair of fields $\theta_a(x)$ and $\phi_a(x)$
defined on the interval $x\in[0,L]$ by 
\begin{equation}
\theta_a(x)=\frac{\vartheta_{a}(x)+\vartheta_{a}(-x)}{2},\quad\phi_a(x)=\frac{\vartheta_{a}(x)-\vartheta_{a}(-x)}{2}
\end{equation}
which are similar to the corresponding fields defined in bosonization of systems with periodic boundary conditions. 
The fields $\theta_a(x)$ and $\phi_a(x)$ are periodic and anti-periodic respectively and 
have the commutation relations $[\theta_a(x),\theta_b(y)]=[\phi_a(x),\phi_b(y)]=0$ and 
$[\theta_a(x),\phi_a(x')]=i\pi[\textrm{sgn}(x-x')-1+\frac{2 x'}{L}]$
so that $\pi_a(x)=\partial_x\phi_a(x)$, which is the density fluctuation 
operator ($\rho_{a,R}(x)+\rho_{a L}(x)$)
 can be considered to be canonically conjugate to the variable
 $\theta_a(x)$ so that  
\begin{equation}
[\pi_a(x),\theta_b(y)]= i\pi[\delta(x-y)-\frac{1}{L}]\delta_{ab}.
\end{equation}
 The $1/L$ term arises from the absence of $k=0$ modes of $\pi_a(x)$ and $\theta_b(x)$.
The Cooper pair operator is given by 
$\psi_{a R}^\dagger(x)\psi_{a L}^\dagger(x)\propto e^{2 i\theta_a(x)}$, so that
 $2 \theta_a(x)$ is related to the fluctuating superconducting phase. 
The transformation of the Hamiltonian to a bosonic 
representation in terms of $\theta_a(x)$ transforms the combination of term $H_0+H_1$ in the Hamiltonian into an exactly
 solvable quadratic form
\begin{equation}
H_0+H_1=\int dx \sum_{a,b}U_{ab}\pi_a\pi_b+V_{ab}\partial\theta_a\partial\theta_b
\end{equation}
where $U_{ab}=v_{F,a}\delta_{ab}+V_{R,R,a ,b }-V_{R,L,a ,b }$ and $V_{ab}=v_{F,a}\delta_{ab}+V_{R,R,a b }+V_{R,L, a ,b }$.
The Cooper pair scattering term $H_2$ in the bosonic representation
 takes the form 
\begin{equation}
H_2\sim \sum_{a\neq b}\int dx V^{(pair)}_{a,b}\cos{\{2(\theta_a(x)-\theta_b(x))\}},
\end{equation}
which, for attractive interactions ($ V^{(pair)}_{a,b}<0$), tends to align the phase variables $\theta_a(x)$ between channels. Note that in realistic 
systems, the spin-orbit coupling is critical for generating the Cooper 
pair scattering term $V^{(pair)}_{a,b}$\cite{Zhang-pwave} and the 
Zeeman splitting controls the number of channels $N_{ch}$. 

\section{Topological superfluid order}\label{seciii}
 In one-dimension, weak attractive interactions between fermions do not
 lead to a long-ranged superfluid order parameter.
 However the Cooper pair tunneling term $H_2$,
becomes relevant  under the renormalization group (RG)
 for attractive interactions $H_1$
so that the phase differences $\theta_a(x)-\theta_b(x)$  develop long-range order  $\expect{\theta_a(x)-\theta_b(x)}=0$ or $\pi$.
 To see this, we estimate the scaling dimension of the Cooper pair tunneling 
operator $P_{ab}(x)=e^{2 i (\theta_a(x)-\theta_b(x))}$, that is contained in $H_2$. The scaling dimension of $P_{ab}(x)$ 
is estimated by calculating the bulk correlator\cite{giamarchi} $\expect{P_{ab}(x)P^*_{ab}(y)}$ 
 using periodic boundary conditions and the unperturbed Hamiltonian $H_0+H_1$ which is quadratic in the fields $\theta_a(x)$.
 The correlators can thus be found using Gaussian integration to be
 $\expect{P_{ab}(x)P_{ab}^*(y)}=\textrm{exp}(\int dk \sum_{\alpha,\beta}\zeta^{*}_\alpha\expect{\theta_\alpha(k)\theta^*_\beta(k)}\zeta_\beta)$ 
where we have used  $\zeta_\alpha=2 e^{i k x}\delta_{a,\alpha}-2 e^{i k y}\delta_{b,\alpha}$.
Substituting, and integrating over $k$ we get $\expect{P_{ab}(0)P_{ab}^*(x)}\sim e^{-\alpha\zeta^T \zeta\log{x} }\sim|x|^{-2\alpha}$
where we have used the matrix-vector notation  $\vec{\zeta}=\zeta_c=2\delta_{a,c}-2\delta_{b,c}$ and 
\begin{equation}
\alpha=\frac{\zeta^T (U^{-1/2}V U^{-1/2})^{-1/2}\zeta}{\zeta^T\zeta}<1
\end{equation} 
provided that $\zeta^T (V-U) \zeta= 2\zeta^T V_{R ,L}\zeta < 0$, \textit{i.e.}, the interaction $V_{R,L}$ is attractive in all channels.
 Thus the Cooper-pair 
operator $P_{ab}(x)$ has a scaling dimension $-\alpha$, with $\alpha<1$, so that $V^{(pair)}_{a,b}$
 is relevant under RG and flows to large values at 
long length-scales. The phase differences  $\theta_a(x)-\theta_b(x)$ develop long-range order
 (\textit{i.e.}, $\expect{\theta_a(x)-\theta_b(x)}=\pi(n_a-n_b)$ for integers $n_a,n_b=0,1$),
so that  
\begin{equation}
\theta_a(x)=n_a\pi+\theta(x)+\delta\theta_a(x)\label{separation}
\end{equation} 
where  $\delta\theta_a(x)\ll \pi$ over long length-scales.

\subsection{Ising pseudo-spin representation}\label{ising}
 The addition of a single fermion $\psi_a^\dagger(x)$, which contains a factor $e^{i\phi_a(x)}$ (Eq.~\ref{bosonization}),
 introduces a kink in the phase $\theta_a(x)$ so that $\theta_a(x)-\theta_b(x)$ changes by $\pi$.
Such a kink costs a finite energy leading to a bulk quasiparticle gap in the superfluid. 
 However, there is no gap to tunneling fermions at the ends since these do not introduce kinks. 
To describe fermion tunneling via such kinks it is necessary to accomodate a position dependent $n_a(x)$ 
for $a=1,\dots,N_{ch}$. Furthermore, the fermion tunneling operators contain fields  $\phi_a(x)$ and $\Delta N_a$  
which are canonically conjugate to $n_a(x)$.

Despite the bulk quasiparticle gap, the ground state of $H$ is degenerate 
 in the bosonic Hilbert space spanned by the operators $\theta_a(x)$,
  since the Hamiltonian $H$ is independent  of the value $n_a=0$ or $1$ for each channel $a$. 
As discussed in the introduction, this is similar to the $Z_2$ order parameter of the one-dimensional Ising model 
with a weak transverse field, where the system is characterized by 
 an order parameter degeneracy (which is $n_a$ in this case) which takes 2 values.
This correspondence is made more explicit by defining the local
 spin-fields for each channel   
\begin{equation}
 S_z^{(a)}(x)=e^{i \pi n_a(x)}\label{Sz}.
\end{equation}
The TS wire in this paper is in an ordered Ising phase $\expect{S_z^{(a)}(x)}=\pm 1$. Furthermore,
 the wire has a $Z_2$ Ising degeneracy for each of the $N_{ch}$ channels.
 Furthermore, the low-energy effective Hamiltonian $H$
 (where spatial variations $\delta\theta_a(x)$ have been integrated out), can be separated 
\begin{equation}
H[\{S_z^{(a}(x)\},\theta(x)]=H_{\theta}[\theta(x)]+H_{ps}[\{S_z^{(a)}(x)\}]\label{Hps}
\end{equation}
where $a=1,\dots,N_{ch}$ and $H_{\theta}[\theta(x)]$\
 arises from the common phase variable
 $\theta(x)$ and keeps track of the global charging energy,
 while $H_{ps}$ is the pseudo-spin ($ps$) Hamiltonian keeping track of the
 degenerate sector described by the integers $n_a=0$ or $1$, which
 we will  map to a pseudo-spin degree of freedom in Sec.~\ref{sectop}.
The Hamiltonian $H_{\theta}[\theta(x)]$ describes massless phase fluctuations of $\theta(x)$ and therefore corresponds to a low-energy effective 
action 
\begin{equation}
S_\theta=\frac{K}{2\pi}\int dx d\tau\left[\frac{(\partial_\tau \theta(x,\tau))^2}{u}+u (\partial_x \theta(x,\tau))^2\right]\label{Stheta}
\end{equation}
where $K$ is the Luttinger parameter associated with the 
phase variable $\theta(x)$ and $u$ is the corresponding phonon velocity.

 The one-dimensional transverse field Ising model is related to the mean-field BCS Hamiltonian of 
a topological spinless $p$-wave wire\cite{kitaev} by a  Jordan-Wigner transformation.
 The order parameter degeneracy of the 
Ising model is mapped to the topological degeneracy of the $p$-wave
wire by the non-local Jordan-Wigner transformation. For our system, 
the bosonization transformation can be thought of as a generalization 
of the non-local Jordan-Wigner transformation that allows us to 
describe the topological degeneracy in terms of an order-parameter 
degeneracy. However, it is not \textit{a priori} obvious that the order-parameter 
degeneracy associated with the $n_a$ integers in our system is 
robust to all local fermionic perturbations. We will establish that 
this is essentially true in Sec.~\ref{secv} and the following sections.

\section{Fixed number Hilbert space}\label{hilbert}
The physical Hilbert space of the nanowire system extends only in a
 subspace of the bosonic Hilbert space with a fixed number of fermions 
 $N_{tot}$. Moreover, the many-body wave-function is $2\pi$ periodic 
in terms of $\theta_a(x)$ and not just $\theta(x)$.
 Thus, the many-body wave-function $\Psi$, when written in terms 
of the variables $n_a$ labeling the ground state sector and the average phase variable
 $\bar{\theta}=\int \frac{d x}{L}\theta(x)$, has the symmetry
\begin{equation}
\Psi[\{n_a+1\},\bar{\theta}+\pi]=\Psi[\{n_a\},\bar{\theta}]\label{symmetry},
\end{equation}
which means that the wave function is
unchanged if we change $\bar{\theta}$ by $\pi$ and simultaneously change
all the $n_a$ by 1 for $a=1,\dots,N_{ch}$.
Since the average phase $\bar{\theta}$ is canonically conjugate to the total fermion number $N_{tot}$, 
the wave-function $\Psi$ has the form 
\begin{equation}
\Psi_{N_{tot}}[\{n_a\},\bar{\theta}]= e^{i N_{tot}\bar{\theta}}\Psi_{N_{tot}}[\{n_a\}]\label{constraint}.
\end{equation}
The symmetry condition Eq.~\ref{symmetry}  reduces to 
$\Psi_{N_{tot}}[\{n_a+1\}]=(-1)^{N_{tot}}\Psi_{N_{tot}}[\{n_a\}]$,
 where  $(-1)^{N_{tot}}$ is the fermion parity.
Furthermore, since the operator $(-1)^{\Delta N_a}=e^{i\pi\Delta N_a}$ 
has the effect of transforming $n_a\rightarrow n_a+1$ (note $[\Delta N_a,n_a]=i$), the number 
conservation constraint has the simple form
\begin{equation}
(-1)^{\sum_a \Delta N_a}=(-1)^{N_{tot}}\label{opconstraint}.
\end{equation}
 The Hilbert 
space constraint from Eq.~\ref{opconstraint} is simply that the 
physical states must be eigenstates of the operators
 $\sum_a\Delta N_a$ from the pseudo-spin space and $N_{tot}$  
with the same eigenvalue parity.

\section{Local end perturbations}\label{secv}
As discussed previously, a degeneracy can be called topological only if 
it is insensitive to all local perturbations. So far, by bosonizing 
the Hamiltonian, we have identified an order parameter (\textit{i.e.}, the 
integers $n_a$) in the bosonic space which may potentially lead to 
a topologically protected degeneracy in the fermionic representation.
However, it is not \textit{a priori} obvious that the bosonic order parameters 
$n_a$ are insensitive to any local perturbation in the fermionic 
representation.
For example, impurity scattering, which breaks momentum conservation, can introduce new 
terms in the Hamiltonian, such as 
 $\psi^\dagger_{a,r}(x)\psi_{b,r'}(x)$. These terms are sensitive to the local value of
 the integer $(n_a-n_b)$ associated with the ground state. 
 The fermion operator $\psi^\dagger$ introduces a kink in 
the phase variable $\theta_a(x)$ that in general creates an orthogonal excited state which is gapped by 
the kink energy. Therefore kinks near the ends of the wire that arise from end scattering are the dominant scattering 
terms that control
the topological degeneracy in a wire with an odd number of channels. However, as we shall see in Sec.~\ref{secix}, in a 
cluster of coupled odd-channel wires, if the number of coupled wires is even in some region of space, bulk scattering can lift 
the topological degeneracy in a subtle way, despite being disfavored by the superconducting energy gap. We shall see that topological 
protection of the degeneracy
requires having a wire with a large number of channels in the bundle of wires.

In contrast to bulk scattering, inter-channel scattering at the ends can have a large effect
in general. This is because the end fermion creation operators
 involve the $\phi_a(x)$ at the ends which vanish and therefore do not create any kinks in the 
phase differences $\theta_a(x)-\theta_b(x)$. In the phase representation, the end scattering terms become    
\begin{align}
&H_{scattering}(x\sim 0)\sim i F_a F_b\cos{(\theta_a(0)-\theta_b(0))}\nonumber\\
&\sim i F_a F_b e^{i\pi (n_a-n_b)}
\end{align} 
at $x=0$ and 
\begin{align}
&H_{scattering}(x\sim L)\sim i F_a F_b e^{i\pi (\Delta N_a-\Delta N_b)}\cos{(\theta_a(L)-\theta_b(L))}\nonumber\\
&\sim i F_a F_b  e^{i\pi (\Delta N_a-\Delta N_b)} e^{i\pi (n_a-n_b)}
\end{align}
at $x=L$.

\section{Topological degeneracy}\label{sectop}
The degeneracy associated with $n_a=0,1$ is in general lowered by
 $H_{scattering}(x\sim 0,L)$. Only degeneracy that survives all weak local 
perturbations qualifies as being topologically protected. 
For simplicity of discussion, we will ignore the 
bulk scattering disorder,
which we have shown does not affect wires with an odd number of channels.
As defined in Sec.~\ref{ising}, the low-energy effective Hamiltonian in the degeneracy space spanned by $n_a$ is conveniently
 defined in terms of the pseudo-spin matrices 
 $S_z^{(a)}=e^{i n_a\pi}$  and  $S_x^{(a)}=e^{i\pi \Delta N_a}$, where we have assumed $n_a(x)=n_a$ to be free of kinks and 
independent of $x$ in Eq.~\ref{Sz}. This is justified for the ground-state of a Hamiltonian of the form in Eq.~\ref{Hps} where there 
is no term creating kinks in the order parameter $n_a(x)$.
 The Klein-factors can be written in terms of spin operators as $F_a=\prod_{b<a}S_x^{(b)}$. 
The scattering terms in the spin-representation are written as $H_{scattering}(x\sim 0)\sim S_z^{(a)}S_z^{(b)}\prod_{a\leq c<b}S_x^{(c)}$, and 
$H_{scattering}(x\sim L)\propto S_z^{(a)}S_z^{(b)}\prod_{a< c\leq b}S_x^{(c)}$.
It follows from Eq.~\ref{opconstraint} that the total fermion-parity operator  can be written as 
\begin{equation}
Q=\prod_b S_x^{(b)},\label{eqq}
\end{equation}
which commutes with all terms in the Hamiltonian.
Thus the eigenstates of the Hamiltonian can be split into 2 sets of even $(Q=1)$ and odd $(Q=-1)$ number of fermions. 
Topological degeneracy of even and odd fermions, similar to that associated with MFs, occurs if there 
exists an operator $\Lambda$ that commutes with all terms in the Hamiltonian $H$ and anti-commutes with the 
fermion parity operator $Q$. In this case 
\begin{align}
&\ket{Q=-1}=\Lambda\ket{Q=1}\\
&E(Q=-1)\ket{Q=-1}=H\ket{Q=-1}=H\Lambda\ket{Q=1}\nonumber\\
&=\Lambda H\ket{Q=1}=\Lambda E(Q=1)\ket{Q=1}=E(Q=1)\ket{Q=-1}
\end{align}
proving that 
\begin{equation}
E(Q=1)=E(Q=-1).
\end{equation}
The operator $\Lambda$ cannot exist in general in the even number of bands case. To see this observe 
that if $\Lambda$ commutes with both the end boundary terms 
 $H_{scattering}(x\sim 0,L)$, it must also commute with the 
product of these operators $S_x^{(a)}S_x^{(b)}$. For even number of bands,
 the total fermion-parity operator $Q$ is a product of such 
operators. Therefore $\Lambda$ commutes with $Q$ as well and cannot anti-commute with $Q$. 

On the other hand, for odd number of bands, $Q$ is a product of an
 odd number of $S_x^a$ and therefore $\Lambda$
 does not in general commute with $Q$.
 In fact, the operator $\Lambda$ can be explicitly constructed as:
\begin{equation}
\Lambda=\prod_a S_y^{(a)}\prod_b S_x^{(2 b+1)}
\end{equation}
which anticommutes with $Q$ since $a$ takes an odd number of values.
 Also $\Lambda$ commutes with both 
boundary scattering terms in $H$.
 Thus $H$ has a topologically protected 
degeneracy between even and odd number fermion modes only in wires with an odd number of channels. 
From here onwards we refer to such wires as topological wires. 
This proof is very similar 
to the proof of the Lieb-Shultz-Mattis theorem by
 Altman and Auerbach \cite{auerbach}.
 Moreover, this result is 
robust to bulk scattering (Eq.~\ref{eq:bulkscattering})
 since this only effects even number of channels.   
However, in the physical Hilbert space determined by the
 constraint Eq.~\ref{opconstraint}, which in the 
spin-representation requires that $Q=(-1)^{N_{tot}}$  where $N_{tot}$ 
is the total number of fermions, the degeneracy sub-spaces $Q=\pm 1$ must have different
 energies because of $H_\theta[N_{tot}]$. This lifts the topological degeneracy even for an isolated wire 
with an odd number of channels.

\begin{figure}
\centering
\includegraphics[scale=0.3,angle=270]{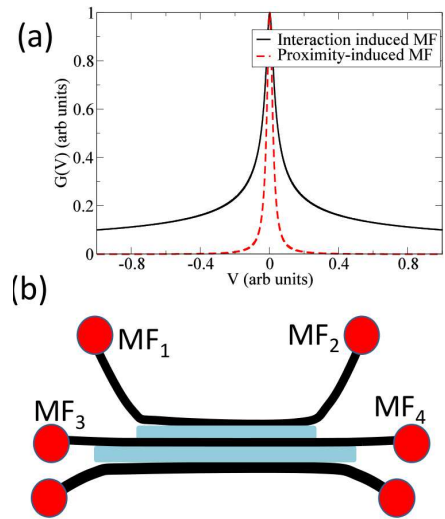}
\caption{(Color online) (a) Tunneling conductance $G(V)$ into 
a Majorana site at the end of a long wire, shows power-law peak
(solid line) characteristic of tunneling into Luttinger liquid with 
attractive interactions. This differs qualitatively from the 
 $\delta$-function peak (dashed line)
 expected from a proximity-coupled 
nanowire.\cite{long-PRB}
 For the plot, we have chosen a 
Luttinger parameter $K=2$ and added a lead-induced lorentzian 
broadening ($(\omega^2+\Gamma^2)^{-1}$) with width  $\Gamma=0.02$.
(b) Schematic figure to obtain a topological degeneracy using 
topological superfluid wires containing phase fluctuations and phase-slips.
Three topological wires (shown in black) each of which
 support MFs (discs at ends) are coupled 
by tunneling (shown as shaded regions) to obtain a topological 
degeneracy which is robust to bulk scattering and respects number conservation.
 }\label{Fig2}
\end{figure}

\section{Topological degeneracy in multiple wires}\label{secxi}
 The topological degeneracy in odd-channel wires is
 broken by the global phase fluctuations which are needed to
 enforce number conservation. 
This makes an isolated wire unusable as a topological qubit.
 The number conservation constraint (Eq.~\ref{constraint}) 
can be relaxed by using a system of three odd channel 
topological wires which are connected in the middle by tunneling 
(Fig.~\ref{Fig2}(b)). The third wire is simply 
to keep the number of channels odd in the contact region so as 
to avoid bulk scattering (Eq.~\ref{eq:bulkscattering}). 
 For simplicity 
of discussion we will focus on the other two wires with end 
MFs $MF_{1,2,3,4}$. Considering 
specifically the case where the two wires have different fermi
 wave-vectors, so that single-particle tunneling between 
the wires is avoided, the dominant tunneling process between the
 wires is Cooper pair tunneling.
The overlapping segment of wire in the middle 
 $[\frac{L}{2}-L_1,\frac{L}{2}+L_1]$ can be taken to have a common 
global phase $\theta(x)$ so that one can define pseudo-spin variables
 $n_a^{(j)}$ for each wire $j=1,2$.
The geometry in Fig.~\ref{Fig2}(b) is such that there is no direct
 tunneling between MFs in the two different wires.
Therefore even though the relative phases of the wires fluctuate away
 from the overlap region, 
 the tunneling Hamiltonian in each wire is unaffected and can be
 written as
  $H=H_{\theta}[N_{tot}]+H_{ps}[n_a^{(1)}]+H_{ps}[n_a^{(2)}]$
 where the constraint on $N_{tot}$ is now related to the total
 fermion parity $Q=Q_1 Q_2$.
 Moreover, the fermion parity operators $Q_{1,2}$ of individual
 wires are decoupled and 
 the states with the same total fermion parity
 $(Q_1,Q_2)=(-1,1)$ and $(Q_1,Q_2)=(1,-1)$ are exactly
 degenerate up to exponential splittings from instanton effects.

\section{Exponential splitting}\label{exponential}
The pseudo-spin Hamiltonian $H_{ps}$ which leads to topological 
degeneracy ignores virtual processes that overcome the tunnel 
barrier to relative phase fluctuations imposed by the term 
 $V^{(pair)}_{ab}\int \cos{2\{\theta_a(x)-\theta_b(x)\}}$. 
Such processes lead to an instanton tunneling\cite{altland} between 
the states with different values of the local order-parameter $n_a-n_b=\pm 1$ in the Ising representation,
 and can be treated as a  
tunneling of the $k=0$ mode of $\theta_a(x)-\theta_b(x)$ from one energy minimum to another\cite{altland}.
The tunneling-induced splitting of the topological degeneracy 
can then be calculated by considering 
the Hamiltonian of the $k=0$ mode of the bosonized Hamiltonian $H=H_0+H_1+H_2$ 
and is found to scale as \cite{golubev}
 $\delta\sim e^{-\pi L\sqrt{V^{(pair)}_{ab}/U_{a b}}}$.

 Similar splitting of the degeneracy occurs from phase slips between
 different wires which would occur in the limit of two proximity-induced
 TS nanowires, with a weak superconducting  link between the two
 associated bulk superconductors. Again, we have energy splittings 
 induced by coherent phase slips between the superconductors, and
 the energy splitting should decrease exponentially with the square-root
 of the ratio between the Josephson coupling and the charging energy
 of the two superconductors.  For bulk superconductors, the latter
 should be determined by the Coulomb charging energy, rather than
 the Fermi energy divided by the wire length or volume.

\section{Local potentials and phase-slips}\label{secix}
In Sec.~\ref{secv}, we argued that the effect of local 
potentials was dominated by scattering at the ends of the wire.
Therefore, so far, we have established a topological degeneracy that is 
robust to local perturbations at the ends of the system.
In this section, we examine more closely the effect of local potentials in 
the bulk of the wire.  
 Such local potential terms of the form  $g\psi_a^\dagger(x)\psi_b(x)$
 introduce kinks in the order parameters $n_a(x),n_b(x)$, with coupling constant $g$,
 that in general leads to an excited state which is gapped by 
the kink energy. The kink generated by such a fermion tunneling term 
in the general case, is associated with an excitation energy that takes
 us outside the ground-state manifold and therefore can only participate 
in virtual excitations that generate low-energy effective Hamiltonian terms which are products of 
tunneling terms. The only such low-energy effective term that can be generated by bulk scattering  is 
\begin{equation}
H_{scattering}^{(bulk)}(x)\sim g^{N_{ch}/2}(-1)^{\sum_a n_a(x)}e^{i\phi(x)}\label{eq:bulkscattering}
\end{equation}
  that generates a kink in all the channels $a$ so that no relative phase difference is created and the 
 resulting wave-function can be described as a kink in the field
 $\theta(x)$. The term $e^{i\phi(x)}$ in Eq.~\ref{eq:bulkscattering} is a vortex creation operator 
in the bosonic phase field $\theta(x)$. The corresponding term in the action, after the phase degree of 
freedom has been integrated out,\cite{kt} can be written as 
  \begin{align}
&S_{scattering}^{(bulk)}(x)\sim g^{N_{ch}/2}(-1)^{\sum_a n_a}e^{K\log {L/\Lambda}}\nonumber\\
&=\left(\frac{g^{1/2}\Lambda^{\kappa}}{L^{\kappa}}\right)^{N_{ch}}(-1)^{\sum_a n_a(x)}
\end{align} 
where the Luttinger parameter $K$ defined in Eq.~\ref{Stheta}, 
 and $\kappa=K/N_{ch}>1$ is a number of order unity.
Such a term, being a product of $N_{ch}/2$ tunneling terms is exponentially small in the number of channels.
 Moreover, since each tunneling term has a pair of fermion operators, such a term can only be
 generated in a nanowire with an even number of channels. The bulk scattering term couples the
 pseudo-spin degree of freedom $n_a(x)$ to the gapless phase modes $\phi(x)$
 making the even channel case useless for TQC. Such bulk
 scattering is avoided by introducing a wire with a very large 
number of channels, as shown in Fig.~\ref{Fig2}(b), so that the
 product term Eq.~\ref{eq:bulkscattering} is exponentially small.
Thus, in order to avoid the bulk scattering, it is
 necessary to couple the nanowire to a superconductor with a large 
number of channels. This effectively requires the introduction
 of a bulk-like superconductor as in the case of the proximity effect.

The bulk-scattering term in Eq.~\ref{eq:bulkscattering}
 can be thought of as an impurity generated quantum phase-slip.
The effects of a phase slip center can be understood, alternatively,
 by starting from the limit of strong backscattering, where the
 center effectively divides the wire bundle  into two almost-separate
 halves.  If the total  number of channels in the wire bundle is odd,
 then dividing the bundle in two pieces leads to a new Majorana state
 on each side of the split, with a small coupling between the two
 Majoranas due to the residual tunneling.  This coupling leads to a
 finite energy splitting between states where the resulting fermion
 state of the pair is occupied or not.  If one introduces a change of
 $2 \pi$ in the phase of the superconductors on one side of the split,
 this will change the sign of the energy splitting, thus  interchanging
 the upper and lower energy states.\cite{kitaev,yakovenko,fu_prl'08,roman,oreg} 
 If the phase slip occurs rapidly,  the fermion number of the
 pair cannot change, so the energies before and after the phase slip
 will differ by the energy splitting between the two states.
  Thus, an isolated phase slip can only occur as a virtual process.
  (A double phase slip is allowed, however. Also,  at finite temperatures, 
single $2 \pi$ phase slips can occur, as  a fermion at the center can
 be further excited into a bulk quasiparticle which is then eventually
 transferred to a low-energy state at the wire end.) If the Josephson
 coupling across the phase slip center is increased, the energy
 splitting between the occupied and unoccupied localized fermion state
 might become equal to the bulk energy gap, which only increases the
 action barrier to forming a single phase slip.

 For a wire with an odd number of channels, quantum tunneling of a
 single $2\pi$ phase slip is not allowed, and phase slips occur only
 in pairs, which have no net effect on the Majorana states.  However,
 when an even number of odd-channel wires is bound together to form
 a bundle with an even-number of channels, backscattering near the
 center of the bundle can lead to a $2 \pi$ quantum phase slip.
  (Fig.~\ref{Fig2}(b))  This has the effect of multiplying the Majorana
 operators to the left of the phase slip by a factor of -1 relative to
 operators on the right.   The cumulative effect of such phase slips is
 to cause oscillations in any operator formed  from a product of odd
 numbers of Majorana operators on opposite sides of the phase-slip
 center, and which signifies an energy splitting  between various states
 in the Majorana fermion Hilbert space.
 [As an alternate description, we may say that the coherent generation
 of phase slips leads to a disruption of the continuous exchange of
 Cooper pairs across the phase-slip center, which  leads in turn to a
 sensitivity of  energy to the total electron parity of  on one side of
 the center.] Similarly, backscattering near a junction of several
 odd-channeled wires can cause a $2 \pi$ phase slip in one pair of
 wires, which would introduce a term in the energy sensitive to the
 total electron parity of the pair.  Consistent with the bosonization results,
 we note  that  
when the number of channels in each wire is large, the rate of
 generation of quantum  phase slips will decrease exponentially with
 the number of channels, for a fixed small amplitude of single-particle
 backscattering,  so that the energy splitting that is non-exponential
 in $L$ will have an amplitude that exponentially small in $N_{ch}$.
  In this way, we find that for a set of sufficiently thick wires with
  large but finite $L$, one can recover the exponential in $L$ behavior
 found in the case of an  adiabatic junction.

We may ask further what happens if the tunnel coupling across a bulk phase slip
 center is taken all the way to zero.  Suppose there is an odd number
 $N_w$  of odd-channel wires in the wire bundle, with each wire having
 a Majorana state at each of its free ends. Including the single
 Majorana state at the phase slip center on each half of the wire
 bundle, there will be an even number $N_w + 1$ of Majorana states on
 each half of the system.  Each wire will then have a number
 $M = 2^{(N_w-1)/2}$  of degenerate ground states, so the total  system
 will have a degeneracy $M^2 = 2^{N_w - 1}$.  When coupling is
 reintroduced, this number stays the same.  The pair of Majoranas near
 the phase slip center are removed from consideration as they now form
 a fermion state with finite energy.  The remaining $2 N_w$ Majorana
 locations now belong to a single wire bundle, with one overall conserved
 charge, so the number of degenerate states is again  $2^{N_w - 1}$.
  However, if the tunnel coupling across the center is very weak, the
 extra  finite fermion state  may have an energy which is well below the
 gap. If the total number of electrons in the system is specified, say,
 to be an even number,  the energy splitting between the two fermion
 states may be thought of as arising from a difference between a state
 of the wire with a small excess electron probability near the center 
(and a small depletion near the ends) compared with one with a small
 depletion near the center.

In the case where $N_w$ is even, there are no extra Majorana states near
 the phase slip center when the two halves are decoupled.  The total
 number of zero-energy states is $2^{N_w - 2}$ when the wires are
 completely decoupled, while it would be $2^{N_w - 1 }$  if there were
 no backscattering and no phase slips allowed.  For the case of weak
 tunneling there would be two sets of  states with $2^{N_w - 2}$
 degenerate states each, but a small energy splitting between the two
 sets.  This energy splitting corresponds to a splitting between states
 where the fermion parity on, say, the right hand side is predominantly
 odd and where it is predominantly even.  (The fermion parity on the
 other side is accordingly determined since  we are considering a total
 system with a definite number of electrons.)

\section{End MFs and tunneling spectrum}\label{secx}
 Even though the exact topological degeneracy may be  
broken by number conservation for single wires, the degeneracy
for odd-number-of-channel wires is only split to order $L^{-1}$ and it should continue to manifest itself as
  a  near-zero energy peak in end-tunneling conductance, as has
 been predicted in previous work. To understand the spatial dependence of the tunneling, it is convenient to 
generalize our definition of the pseudo-spin used in Sec.~\ref{sectop} so that $S_x^{(a)}(x)$ and $S_y^{(a)}(x)$
 are position dependent field operators. The order parameter operator $S_z^{(a)}(x)$ was already defined as a position dependent 
field in Eq.~\ref{Sz}. The generalized $x$ spin operators are defined as $S_x^{(a)}(x)\sim e^{i[\phi_a(x)+\Delta N_a x \pi/L]}$
 using the fact that $\phi_a(x)+\frac{\Delta N_a x}{L}$ 
is canonically conjugate to $\theta_a(x)$, and therefore generates a kink in $n_a(x)$. The corresponding $y$ pseudo-spin operator 
is defined as $S_y^{(a)}(x)=i S_z^{(a)}(x)S_x^{(a)}(x)$.  Using the bosonization identity Eq.~\ref{bosonization} 
in conjunction with the Eq.~\ref{separation}, the fermion tunneling operator at the ends can be decomposed as
  $\psi_a^\dagger(x\sim 0)\sim e^{i\theta(x)}\gamma_{a,R}(x)$
into a manifestly Hermitean (and therefore Majorana part) 
 $\gamma_{a,R}^\dagger(x\sim L)\sim F_a S_y^{(a)}(x)$ 
 and a phase fluctuation operator $e^{i\theta(x)}$. Similarly, the left fermion tunneling operator
  $\psi_a^\dagger(x\sim L)\sim e^{i\theta(x)}\gamma_{a,L}(x)$
can be shown to contain a Majorana part $\gamma_{a,L}^\dagger(x\sim 0)\sim F_a S_z^{(a)}(x)$.
   The operators $S_x^{(a)}(x)$ and $S_y^{(a)}(x)$ are localized at the ends of the wire 
since the bulk of the $S_z^{(a)}(x)$ ordered Ising phases are gapped. Moreover, the Hamiltonian 
$H_{ps}[n_a]$ in the pseudo-spin representation and the
 corresponding eigenstates $\ket{Q=\pm 1}$ are both real. Therefore in this 
eigenbasis, the left and right Majorana operators are given by $\gamma_L=\sigma_x$ and $\gamma_R=\sigma_y$ respectively, where
$\sigma_{x,y}$ are the Pauli matrices in the pseudo-spin basis where $\ket{Q=\pm 1}$ are eigenstates of $\sigma_z$.
While tunneling into these idealized MF modes would have generated a sharp zero-energy peak similar to the proximity-induced
 case\cite{long-PRB},  the end tunneling conductance into the physical fermions is dominated by the phase mode $\theta(x)$ and is 
essentially identical to tunneling into the end of a single spin-less fermion channel with
 attractive interactions \cite{gogolin,giamarchi}: 
\begin{equation}
G(V)\propto V^{-\frac{K-1}{ K}} 
\end{equation}
where $K>1$, so that there is a power-law divergence
 at $V\rightarrow 0$ (Fig.~\ref{Fig2}(a)),
 instead of the $\delta(V)$ from the proximity effect 
theory without phase fluctuations.

More interestingly, for a finite wire of length $L$, the phonon spectrum is gapped with energy levels
 given by $\omega_n=(n+1)\hbar u/L$, where $u$ is the phonon velocity defined in Eq.~\ref{Stheta}. Such a 
finite wire leads to a discretized spectral function (instead of the continuum one shown in Fig.~\ref{Fig2}(a))
 consisting of a sequence of narrow peaks reflecting the phonon frequencies. One can also consider the possibility 
of tunneling a single fermion into the Majorana end state from a neighboring confining potential such as
 a quantum dot \cite{flensberg}. Coupling the Majorana end state to such a localized fermion state not only switches the 
fermion parity $Q$ but also excites a wave-packet of phonons that propagates from the end. Such a phonon packet can only be avoided 
 by performing the tunneling operation on a time-scale that is adiabatic compared to the inverse lowest phonon frequency 
$\omega_0^{-1}\sim L/\hbar c$ which scales with the length of the system.

\section{Topologically protected operations}\label{secxii}
In earlier sections we have shown that a pair of topological wires
 have a topological degeneracy which is only split by an amount that decreases 
rapidly with the length of the wire $L$ and therefore may be used as a
 topologically protected qubit. However, the question of whether this topologically 
protected qubit can be used to perform topological quantum computation is not 
obvious. In particular, the presence of phase-fluctuations and phase-slips can 
potentially lead to decoherence of otherwise topologically protected manipulations 
of the qubit. Here, we will restrict our discussion to qubit read-out and MF braid operations, 
which are the 2 classes of topologically protected manipulations that have been proposed for
 topological systems with Majorana fermions.
While not sufficient to perform topological quantum computation by themselves, they are enough 
to verify directly the non-Abelian statistics of MFs.
 Recently several schemes have been proposed for performing the operations of 
read-out of topological qubits \cite{hassler,alicea1,flensberg} and braiding of MFs. 
\cite{alicea1,david,david1,flensberg} The proposed read-out schemes that use either
 interferometry-based fermion parity measurement,
 \cite{hassler} capacitive charge read-out,\cite{alicea1,david1} or tunneling. \cite{david,flensberg}
On the other hand, the braiding schemes require creation and annihilation of tri-junctions in the 
wires.\cite{alicea1} All the read-out operations and the creation and annihilation of tri-junctions 
can in principle be implemented for the topological superfluid wire systems discussed in this paper 
by a straightforward generalization of the corresponding proposals.
However, some care must be taken when implementing these proposals to avoid effects due to phase-fluctuations
 and phase-slips during the read-out and braid operations, which can lead to unwanted decoherence.  The additional
 requirements, together with effects of finite phonon propagation velocities, can lead to limitations on the rate
 at which manipulations can be performed. A full discussion of these issues would require a long digression, 
 which we postpone to a future publication.

\section{Conclusion}
In this paper we have shown how a topological degeneracy emerges
 in a number conserving Hamiltonian where superfluidity is induced
 in a multi-channel nanowire with an odd number of channels by attractive 
interactions. Our treatment uses the bosonization formulation and 
therefore goes beyond the BCS mean-field $+$ fluctuation treatment 
in a systematic way so that number conservation is clearly preserved.
In the bosonization representation, which is related to the fermions 
in a non-local way, the topological order of the fermionic Hamiltonian 
finds a natural interpretation as a $Z_2$ pseudo-spin order of the
 relative phases between various channels.
 However, the apparent degeneracy of $2^{N_{ch}}$,
where $N_{ch}$ is the number of channels, is explicitly shown to 
be broken down by end scattering to $2^{N_{ch}(mod\, 2)}$. Therefore 
we show in the general interacting case that the topological degeneracy 
survives end scattering only for wires with an odd number of channels.
Local perturbations in the middle of the wire introduce quantum
phase-slips that are shown to lift
the topological degeneracy in general by an energy-splitting  that scales
as $g^{N_{ch}/2} L^ {-  \kappa  N_{ch}}$, where $g$ measures the strength of
the impurity potential and $\kappa$ is a constant of order unity. 
 Thus the topological degeneracy which is robust to all local
 perturbation  occurs only for a bundle of   wires where one of
 the wires has a large number of channels.  
  
The number conservation constraint appears in a simple way as a 
constraint between the pseudo-spin and phase sectors of the Hamiltonian.
This constraint in general lifts the topological degeneracy of the 
topological nanowires unless multiple nanowires are considered.
 Despite the $1/L$ splitting of the 
topological degeneracy, an isolated odd channeled nanowire 
 shows  a peak in the tunneling spectrum. Unlike the proximity-induced 
superconducting case, the peak is strongly broadened by
 phase fluctuations.
Despite the presence of phase fluctuations, we
find  nearly perfect degeneracy for the Majorana states in a multiwire
bundle, provided that phase slips are sufficiently suppressed in wire
sections containing an even number of  channels.

As expected, our results approach the mean-field BCS
limit in the case where the topological superfluid wire is
in contact with an infinite bulk superconductor, which
creates an ensemble of number states in the wire. In this
case, a system of wires which has $n$ MFs has a degen-
eracy of $2^{n/2}$. The bosonization approach in this paper
allows us to study how the limit of an infinite system
is approached and explicitly consider the role of global
and local number conservation, which are not manifestly
obeyed by the mean-field formalism. Imposing global
number conservation on the mean-field results splits the
degenerate ground-state manifold into 2 groups of $2^{n/2-1}$
degenerate states with definite number of fermions in
each state. The splitting between the 2 groups of states
vanishes as an inverse power of the system size in the
limit of an large system. Therefore, the splitting in the
topologically degenerate manifold from global number
conservation vanishes in the limit of an infinite system
in agreement with the BCS mean-field limit. The power law dependence on
 system size is too slow to be useful for quantum computation. This is not a problem,
 however, as manipulations such as braiding of Majorana sites can be performed
 within a  $2^{N/2-1}$ dimensional subspace which has definite particle number.   

The power-law dependence of energy splittings due to global number conservation can
 be a serious obstacle to topological computation if at intermediate times the system
 is divided into isolated wire segments with number conservation for the individual segments.
 The resulting multiple energy splittings can then lead to rapid dephasing of states
 in the computational Hilbert space. 

Local number conservation, which is present in our superfluid Hamilto-
nian, manifests itself in more subtle effects such as local phase fluctuations
 and quantum phase slips. While  local
phase fluctuations do not lead to any additional split-
ting of the degeneracy, as discussed in Sec.~\ref{secxi}, phase-slips split the topological degeneracy.
 When the number of channels $N_{ch}$ becomes large, however, these splittings become rapidly small, and
 they are typically much smaller than the spittings due to global number conservation.   For fixed system length,
 we find that the splittings decrease exponentially with $N_{ch}$, while for fixed large $N_{ch}$, the splittings
 decrease as a large inverse power of the system length. 
Therefore, our result approaches the mean-field BCS result in the limit of
an infinite system size and an infinite number of channels,
both of which can be realized in the proximity-induced
superconductivity case from a large bulk superconductor
as in the previous proposals.\cite{roman,oreg}

J.S acknowledges the opportunity to present some of these results at the Gump Station 2011 workshop in Moorea which motivated some of the 
presentation of the material in this paper.
 This work was supported by DARPA-QuEST, NSF grant DMR- 0906475,
 JQI-NSF-PFC, and Microsoft-Q.

\end{document}